\documentclass{article}
\usepackage{amsmath}

\begin{document}

\textbf{THE CONSTITUTIVE RELATIONS AND THE}

\textbf{MAGNETOELECTRIC EFFECT FOR \ MOVING\ MEDIA\bigskip \medskip }

Tomislav Ivezi\'{c}

\textit{Ru%
\mbox
 {\it{d}\hspace{-.15em}\rule[1.25ex]{.2em}{.04ex}\hspace{-.05em}}er Bo\v{s}%
kovi\'{c} Institute, P.O.B. 180, 10002 Zagreb, Croatia}

E-mail: ivezic@irb.hr\textbf{\bigskip \bigskip }

\noindent In this paper the constitutive relations for moving media with
homogeneous and isotropic electric and magnetic properties are presented as
the connections between the generalized magnetization-polarization bivector $%
\mathcal{M}$ and the electromagnetic field $F$. Using the decompositions of $%
F$ and $\mathcal{M}$, it is shown how the polarization vector $P(x)$ and the
magnetization vector $M(x)$ depend on $E$, $B$ and two different velocity
vectors, $u$ - the bulk velocity vector of the medium, and $v$ - the
velocity vector of the observers who measure $E$ and $B$ fields. These
constitutive relations with four-dimensional geometric quantities, which
correctly transform under the Lorentz transformations (LT), are compared
with Minkowski's constitutive relations with the 3-vectors and several
essential differences are pointed out. They are caused by the fact that,
contrary to the general opinion, the usual transformations of the 3-vectors $%
\mathbf{E}$, $\mathbf{B}$, $\mathbf{P}$, $\mathbf{M}$, etc. are not the LT.
The physical explanation is presented for the existence of the
magnetoelectric effect in moving media that essentially differs from the
traditional one.\bigskip

\bigskip \bigskip

\noindent \textbf{1. Introduction\bigskip }

\noindent In this paper the constitutive relations for moving media with
homogeneous and isotropic electric and magnetic properties are presented
using the abstract four-dimensional (4D) geometric quantities and their
representations in the standard basis. These results are compared with
Minkowski's constitutive relations$^{1}$ with the 3-vectors and with
Minkowski's constitutive relations that are obtained by means of exterior
forms.$^{2}$ The paper is organized as follows.

First, in Sec. 2, we present a review of some previous results that are
important for the theory presented here. In the recent paper,$^{3}$ a
formulation of the field equations for moving media is developed by the
generalization of an axiomatic geometric formulation of the electromagnetism
in vacuum.$^{4}$ As mentioned in Ref. 3 almost the entire physics literature
deals with the electromagnetic excitation tensor $\mathcal{H}$, i.e. with
the electric $D$ and magnetic $H$ excitations (see Eq. (\ref{h1})) and the
constitutive relations refer to the connections between them and $F$, i.e. $%
E $ and $B$, respectively. But, as shown in Ref. 3, it is physically better
founded to formulate the field equations for moving media in terms of $F$
and the generalized magnetization-polarization bivector $\mathcal{M}$
instead of, as usual, $F$ and $\mathcal{H}$. In Sec. 2, the decompositions
of $F$ (\ref{E2}) and $\mathcal{M}$ (\ref{M1}) are presented. $F$ is
decomposed into vectors $E$, $B$ and $v$ - the velocity vector of the
observers who measure $E$ and $B$ fields. $\mathcal{M}$ is decomposed into
the polarization vector $P$, the magnetization vector $M$ and $u$ - the bulk
velocity vector of the medium. In Ref. 3, the field equations are written in
terms of $F$ and $\mathcal{M}$ and also in terms of vectors $E$, $B$, $P$, $%
M $ and the velocity vectors $u$ and $v$. These field equations are also
quoted in Sec. 2. Furthermore, in Sec. 2, the usual transformations (UT) of
the 3-vectors $\mathbf{E}$ and $\mathbf{B}$, (\ref{JCB}), and of $\mathbf{P}$
and $\mathbf{M}$, (\ref{ps}), are written and their difference relative to
the Lorentz transformations (LT) of vectors, as 4D geometric quantities,
e.g., the electric field vector $E$, (\ref{T1}), is pointed out.

In Sec. 3, we formulate the constitutive relations as the relations between $%
\mathcal{M}$ and $F$, Eqs. (\ref{cr1}) and (\ref{cr2}). Then, using the
decompositions of $F$ (\ref{E2}) and $\mathcal{M}$ (\ref{M1}) we get from (%
\ref{cr1}) and (\ref{cr2}) how $P(x)$, (\ref{P}), and $M(x)$, (\ref{M}),
depend on $E$, $B$ and two different velocity vectors, $u$ and $v$. The
constitutive relations (\ref{P}) and (\ref{M}) are the basic relations that
are obtained in this paper and they are not reported in previous approaches.
The last term in (\ref{P}) and (\ref{M}) describes the magnetoelectric
effect in a moving dielectric in a new way.

In Sec. 4, we have represented all 4D geometric quantities from (\ref{P})
and (\ref{M}) in the standard basis in order to compare them with some usual
formulations. This procedure yields Eqs. (\ref{po}) and (\ref{ma}).

In Sec. 5, Minkowski's constitutive relations with the 3-vectors (\ref{de})
are quoted. The equations (\ref{de}) are considered to be the fundamental
constitutive equations for moving media in the whole physics community. In
Sec. 5.1, the relations (\ref{de}) are written in equivalent forms as\
constitutive equations which explicitly express the 3-vectors $\mathbf{P}$
and $\mathbf{M}$ as the functions of the 3-vectors $\mathbf{E}$ and $\mathbf{%
B}$, (\ref{pl}) and (\ref{mg}), or (\ref{mp}). These forms of Minkowski's
constitutive relations are compared with our relations (\ref{po}) and (\ref%
{ma}), i.e., with Eqs. (\ref{pc}) and (\ref{ma1}), and several essential
differences are pointed out. It is argued that these differences appear
since Minkowski's constitutive relations are with the 3-vectors and they are
derived using the UT and not the LT. In Sec. 5.2, it is shown that the same
differences remain in the low velocity limit. In Sec. 5.3, it is presented
the comparison with Minkowski's constitutive relations that are obtained by
means of exterior forms.$^{2}$ It is shown that the constitutive relations
with exterior forms from Ref. 2 are completely equivalent to Minkowski's
constitutive relations with the 3-vectors and, accordingly, they also differ
from the relations obtained in this paper.

In Sec. 6, Eq. (\ref{i}) represents the interaction term in the Lagrangian
for the interaction between the electromagnetic field $F$ and the dipole
moment bivector $D$, whereas Eq. (\ref{1}) is its low velocity limit. The
last two terms in (\ref{i}), or (\ref{1}), contain the direct interaction of
$E$ with the magnetic dipole moment vector$\ m$ and $B$ with the electric
dipole moment vector $d$. These terms give the physical explanation for the
existence of the magnetoelectric effect in moving media. That explanation
markedly differs from the traditional one.

In Sec. 7, some remarks are given, which refer to the general constitutive
relations.

In Sec. 8, the conclusions are presented.\bigskip \bigskip

\noindent \textbf{2. A Short Review\ of Some\ Previous Results\bigskip }

\noindent We shall deal with 4D geometric quantities, i.e. in the geometric
algebra formalism. For the exposition of the geometric algebra see Ref. 5.
The generators of the spacetime algebra are four basis vectors $\left\{
\gamma _{\mu }\right\} ,\mu =0...3,$ satisfying $\gamma _{\mu }\cdot \gamma
_{\nu }=\eta _{\mu \nu }=diag(+---)$. This basis, the standard basis, is a
right-handed orthonormal frame of vectors in the Minkowski spacetime $M^{4}$
with $\gamma _{0}$ in the forward light cone, $\gamma _{0}^{2}=1$ and $%
\gamma _{k}^{2}=-1$ ($k=1,2,3$). The standard basis $\left\{ \gamma _{\mu
}\right\} $ corresponds to Einstein's system of coordinates in which the
Einstein synchronization of distant clocks$^{6}$ and Cartesian space
coordinates $x^{i}$ are used in the chosen inertial frame of reference.

First, we briefly expose the formulation of the field equations from Ref. 3.
As shown in Ref. 3, the field equations (5), $\partial (\varepsilon _{0}F+%
\mathcal{M})=j^{(C)}/c$; $\partial \cdot (\varepsilon _{0}F+\mathcal{M}%
)=j^{(C)}/c$, $\partial \wedge F=0$, can be taken as the primary equations
for the electromagnetism in moving media. The bivector $F=F(x)$ represents
the electromagnetic field and, as shown in Ref. 4, it can be taken as the
primary quantity for the whole electromagnetism. $j^{(C)}$ is the conduction
current density of the \emph{free} charges. $\mathcal{M}$ is the generalized
magnetization-polarization bivector $\mathcal{M=M}(x)$, which is connected
with the magnetization-polarization current density of the \emph{bound}
charges $j^{(\mathcal{M})}=-c\partial \mathcal{M}=-c\partial \cdot \mathcal{M%
}$. (According to Eq. (3),$^{3}$ the total current density vector $j$ can be
decomposed as $j=j^{(C)}+j^{(\mathcal{M})}$.) The field equation with
sources is written in the `source representation' in Eq. (7)$^{3}$ $\partial
\cdot \varepsilon _{0}F=j^{(C)}/c-\partial \cdot \mathcal{M}$; the sources
of $F$ are the true currents $j^{(C)}$ and the magnetization-polarization
current density $\partial \cdot \mathcal{M}$. In most materials $\mathcal{M}$
is a function of the field $F$ and this dependence is determined by the
constitutive relations. In all previous formulations of the electromagnetism
in media (at rest, or moving), the electromagnetic excitation tensor is
introduced as $\mathcal{H}=\varepsilon _{0}F+\mathcal{M}$. Then, the
constitutive relations refer to the connections between $\mathcal{H}$ and $F$%
. However, as discussed in Ref. 3, physically different kind of entities are
mixed in that definition for $\mathcal{H}$; an electromagnetic field $F$ and
a matter field $\mathcal{M}$. Furthermore, in general, two different
velocity vectors, $v$ - the velocity of the observers and $u$ - the velocity
of the moving medium, enter into the decompositions of $F$ and $\mathcal{M}$%
, respectively. For $F$, that decomposition is given by Eq. (10)$^{3}$

\begin{equation}
F=E\wedge v/c+(IcB)\cdot v/c,  \label{E2}
\end{equation}%
where the electric and magnetic fields are represented by vectors $E(x)$ and
$B(x)$ and $I$ is the unit pseudoscalar. There is no rest frame for the
field $F$, that is, for $E$ and $B$, and therefore the vector $v$ in the
decomposition (\ref{E2}) is interpreted as the velocity vector of the
observers who measure $E$ and $B$ fields. Then $E(x)$ and $B(x)$ are defined
with respect to $v$, i.e., with respect to the observer, as
\begin{equation}
E=F\cdot v/c,\quad B=-(1/c)I(F\wedge v/c).  \label{E1}
\end{equation}%
It also holds that $E\cdot v=B\cdot v=0$; both $E$ and $B$ are space-like
vectors and they depend not only on $F$ but on $v$ as well. Similarly, in
Eq. (12),$^{3}$ the bivector $\mathcal{M(}x\mathcal{)}$ is decomposed into
two vectors, the polarization vector $P(x)$ and the magnetization vector $%
M(x)$ and the unit time-like vector $u/c$
\begin{equation}
\mathcal{M}=P\wedge u/c+(MI)\cdot u/c^{2}.  \label{M1}
\end{equation}%
There is the rest frame for a medium, i.e., for $\mathcal{M}$, or $P$ and $M$%
, and therefore the vector $u$ in the decomposition (\ref{M1}) is identified
with bulk velocity vector of the medium in spacetime. Then, $P(x)$ and $M(x)$
are defined with respect to $u$ as%
\begin{equation}
P=\mathcal{M}\cdot u/c,\quad M=cI(\mathcal{M}\wedge u/c)  \label{M2}
\end{equation}%
and it holds that $P\cdot u=M\cdot u=0$. As in the case with $F$, it is
visible from (\ref{M2}) that $P$ and $M$ depend not only on $\mathcal{M}$
but on $u$ as well.

Usually, only the velocity vector $u$ of the moving medium is taken into
account, or the case $u=v$ is considered,$^{7}$ i.e., it is supposed that
the observer frame is comoving with medium, or both decompositions (\ref{E2}%
) and (\ref{M1}) are made with the same velocity vector, either $u$ or $v$.$%
^{8}$ Such assumptions enable the introduction of the electromagnetic
excitation bivector $\mathcal{H}$, and, by using (\ref{E2}) and (\ref{M1}),
one finds the decomposition of $\mathcal{H}$ into the electric and magnetic
excitations (other names of which are `electric displacement'\ and `magnetic
field intensity')%
\begin{equation}
\mathcal{H=}D\wedge u/c+(IH)\cdot u/c^{2}  \label{h1}
\end{equation}%
(Eq. (14)$^{3}$), where, as usual, the electric displacement vector $%
D=\varepsilon _{0}E+P$ and the magnetic field intensity vector $H=(1/\mu
_{0})B-M$ are introduced.

In Ref. 3, inserting the decompositions of $F$ and $\mathcal{M}$\ (Eqs. (\ref%
{E2}) and (\ref{M1}) here) into the field equation, the general form of the
field equation for a magnetized and polarized moving medium with $E(x)$, $%
B(x)$, $P(x)$ and $M(x)$ (the Amp\`{e}rian form) is obtained in the form of
Eq. (15), i.e., (16) (the vector part, with sources), or (18) in the `source
representation' $\partial \cdot \{\varepsilon _{0}[E\wedge v/c+(IB)\cdot
v]\}=j^{(C)}/c-\partial \cdot \lbrack P\wedge u/c+(1/c^{2})(MI)\cdot u]$,
and (17) (the trivector part, without sources) $\partial \wedge \lbrack
E\wedge v/c+(IB)\cdot v]=0$. In contrast to all previous results, in the
vector part, i.e., the part with sources, of the field equation there are
two different velocities $u$ and $v$. From these field equations it is
concluded$^{3}$ that in the field equation with sources, (16) or (18) in
Ref. 3, \emph{the usual Amp\`{e}r-Maxwell law and Gauss's law are
inseparably connected in one law.} Similarly, \emph{Faraday's law and the
law that expresses the absence of magnetic charge are also inseparably
connected in one law, }the field equation without sources, Eq. (17).$^{3}$
As shown in Secs. 6 and 7 in Ref. 3, this inseparability is an essential
difference relative to the usual Maxwell equations with the 3-vectors.

Next, we mention an important result regarding the usual formulation of
electromagnetism, as in Ref. 9, which is presented in Ref. 10 and discussed
in Refs. 11, 12 and 3. It is argued$^{10}$ that an individual vector has no
dimension; the dimension is associated with \emph{the dimension of its domain%
}. Hence, the time-dependent $\mathbf{E(r,}t\mathbf{)}$, $\mathbf{B(r,}t%
\mathbf{)}$, $\mathbf{D}(\mathbf{r},t)$ etc. cannot be the 3-vectors, since
they are defined on the spacetime. Therefore, we use the term `vector' for a
geometric quantity, which is defined on the spacetime and which always has
in some basis of that spacetime, e.g., the standard basis $\{\gamma _{\mu
}\} $, four components (some of them can be zero). Note that vectors are
usually called the 4-vectors. However, an incorrect expression, the
3-vector, will still remain for the usual $\mathbf{E(r,}t\mathbf{)}$, $%
\mathbf{B(r,}t\mathbf{)}$, $\mathbf{D}(\mathbf{r},t)$ etc.. Moreover,
recently,$^{13-17,11} $ it is proved that, contrary to the general belief,
the UT of the 3-vectors of the electric and magnetic fields, $\mathbf{E(r,}t%
\mathbf{)}$ and $\mathbf{B(r,}t\mathbf{)}$ respectively, see, e.g., Eqs.
(11.148) and (11.149) in Ref. 9, differ from the LT (boosts) of the
corresponding 4D quantities that represent the electric and magnetic fields.
As explained,$^{13-17,11}$ the fundamental difference between the UT and the
LT of the electric and magnetic fields is that in the UT, e.g., the
components of the transformed $\mathbf{E}^{\prime }$ are expressed by the
mixture of components of $\mathbf{E}$ and $\mathbf{B}$, and similarly for $%
\mathbf{B}^{\prime }$, Eq. (11.148).$^{9}$ The UT of the 3-vectors $\mathbf{E%
}$ and $\mathbf{B}$ are given, e.g., by Eqs. (11.149)$^{9}$ and they are

\begin{eqnarray}
\mathbf{E}^{\prime } &=&\gamma (\mathbf{E}+\mathbf{\beta \times }c\mathbf{B)-%
}(\gamma ^{2}/(1+\gamma ))\mathbf{\beta (\beta \cdot E),}  \notag \\
\mathbf{B}^{\prime } &=&\gamma (\mathbf{B}-(1/c)\mathbf{\beta \times E)-}%
(\gamma ^{2}/(1+\gamma ))\mathbf{\beta (\beta \cdot B),}  \label{JCB}
\end{eqnarray}%
where $\mathbf{E}^{\prime }$, $\mathbf{E}$, $\mathbf{\beta }$ and $\mathbf{B}%
^{\prime }$, $\mathbf{B}$ are all 3-vectors. All what is stated for the
3-vectors $\mathbf{E}$ and $\mathbf{B}$ and their UT holds in the same
measure for the couple of the 3-vectors $\mathbf{P}$ and $\mathbf{M}$\ and
their UT

\begin{eqnarray}
\mathbf{P} &=&\gamma (\mathbf{P}^{\prime }+\mathbf{\beta \times M}^{\prime
}/c\mathbf{)-}(\gamma ^{2}/(1+\gamma ))\mathbf{\beta (\beta \cdot P}^{\prime
}\mathbf{),}  \notag \\
\mathbf{M} &=&\gamma (\mathbf{M}^{\prime }-\mathbf{\beta \times }c\mathbf{P}%
^{\prime }\mathbf{)-}(\gamma ^{2}/(1+\gamma ))\mathbf{\beta (\beta \cdot M}%
^{\prime }\mathbf{),}  \label{ps}
\end{eqnarray}%
see the equations, e.g., Eq. (4.2),$^{18}$ or Eqs. (18-68) - (18-71),$^{19}$
or (6.78a) and (6.81a),$^{20}$ etc.

However, \emph{the correct LT always transform the 4D algebraic object
(vector, bivector) representing the electric field only to the electric
field, and similarly for the magnetic field.}

In order to explain this fundamental difference between the LT and the UT
let us introduce the frame of `fiducial'\ observers as the frame in which
the observers who measure fields $E$ and $B$ are at rest. That frame with
the standard basis $\{\gamma _{\mu }\}$ in it is called the $\gamma _{0}$%
-frame. In the $\gamma _{0}$-frame $v=c\gamma _{0}$ and therefore $E$ from (%
\ref{E1}) becomes $E=F\cdot \gamma _{0}$ and it transforms under the active
LT (Eqs. (10) and (11) in Ref. 11) in such a manner that both $F$ and the
velocity of the observer $v=c\gamma _{0}$ are transformed by the LT, see Eq.
(12)$^{11}$ ($E=F\cdot \gamma _{0}\longrightarrow E^{\prime }=R(F\cdot
\gamma _{0})\widetilde{R}=(RF\widetilde{R})\cdot (R\gamma _{0}\widetilde{R})$%
). As explained in Ref. 11, \emph{Minkowski, in Sec. }11.6,$^{1}$\emph{\
showed that both factors of the vector} $E$, \emph{as the product of one
bivector and one vector, has to be transformed by the LT.} However, it is
worth mentioning that Minkowski in all other parts of Ref. 1 dealt with the
usual 3-vectors $\mathbf{E}$, $\mathbf{B}$, $\mathbf{D}$, etc.. These
correct LT give that
\begin{equation}
E^{\prime }=E+\gamma (E\cdot \beta )\{\gamma _{0}-(\gamma /(1+\gamma ))\beta
\},  \label{T1}
\end{equation}%
Eq. (13).$^{11}$ \emph{In the same way vector }$B$ \emph{transforms and
vectors} $P$, $M$ \emph{as well}, \emph{but for} $P$ \emph{and} $M$ \emph{%
the LT,} \emph{like} (\ref{T1}), \emph{are the transformations from the rest
frame of the medium }($u=c\gamma _{0}$). For boosts in the direction $\gamma
_{1}$ one has to take in that equation that $\beta =\beta \gamma _{1}$ (on
the l.h.s. is vector $\beta $ and on the r.h.s. $\beta $ is a scalar).
Hence, in the standard basis and when $\beta =\beta \gamma _{1}$ that
equation becomes
\begin{equation}
E^{\prime \nu }\gamma _{\nu }=-\beta \gamma E^{1}\gamma _{0}+\gamma
E^{1}\gamma _{1}+E^{2}\gamma _{2}+E^{3}\gamma _{3},  \label{T2}
\end{equation}%
what is Eq. (14).$^{11}$ The most important result is that \emph{the
electric field vector }$E$ \emph{transforms by the LT again to the electric
field vector }$E^{\prime }$\emph{; there is no mixing with the magnetic field%
} $B$. The same happens with vectors $P$ and $M$.

On the other hand, if in the transformation of $E=F\cdot \gamma _{0}$ only $%
F $ is transformed by the LT $R$, but not the velocity of the observer $%
v=c\gamma _{0}$ ($E=F\cdot \gamma _{0}\longrightarrow E_{F}^{\prime }=(RF%
\widetilde{R})\cdot \gamma _{0}$, Eq. (15)$^{11}$), then, in the standard
basis and when $\beta =\beta \gamma _{1}$, one finds

\begin{equation}
E_{F}^{\prime \nu }\gamma _{\nu }=E^{1}\gamma _{1}+\gamma (E^{2}-c\beta
B^{3})\gamma _{2}+\gamma (E^{3}+c\beta B^{2})\gamma _{3}.  \label{J2}
\end{equation}%
what is Eq. (17).$^{11}$ It is visible from the comparison of Eq. (\ref{J2})
with Eq. (11.148)$^{9}$ that the transformations of components (taken in the
standard basis) of $E_{F}^{\prime }$ are exactly the same as the
transformations of $E_{x,y,z}$ from Eq. (11.148),$^{9}$ i.e., the components
from (\ref{JCB}).\bigskip \bigskip

\noindent \textbf{3. Constitutive Relations for Moving Media in Geometric
Terms\bigskip }

\noindent Let us consider the case of a simple medium with homogenous and
isotropic electric and magnetic properties. In accordance with Sec. 2 we
formulate the constitutive relations for moving media using the generalized
magnetization-polarization bivector $\mathcal{M}$ and the electromagnetic
field bivector $F$%
\begin{equation}
\mathcal{M}\cdot u=\varepsilon _{0}\chi _{E}F\cdot u,  \label{cr1}
\end{equation}%
\begin{equation}
(I\mathcal{M})\cdot u=(\chi _{B}/\mu _{0}c^{2})u\cdot (IF)  \label{cr2}
\end{equation}%
and the electric and magnetic susceptibility ($\chi _{E}$, $\chi _{B}$).

Using the decompositions of $F$ (\ref{E2}) and $\mathcal{M}$ (\ref{M1}) we
get from (\ref{cr1}) how the polarization vector $P(x)$ depends on $E$, $B$
and $u$, $v$,%
\begin{equation}
P=(\varepsilon _{0}\chi _{E}/c)\{(1/c)[(u\cdot v)E-(u\cdot E)v]+(u\wedge
v\wedge B)I\}  \label{P}
\end{equation}%
and from (\ref{cr2}) how the magnetization vector $M(x)$ depends on $E$, $B$
and $u$, $v$,%
\begin{equation}
M=\varepsilon _{0}\chi _{B}\{[(u\cdot v)B-(u\cdot B)v]-(1/c)(u\wedge v\wedge
E)I\}.  \label{M}
\end{equation}%
Observe that $P$ in (\ref{P}) and $M$ in (\ref{M}) contain both velocities $%
u $, the bulk velocity vector of the medium, and $v$, the velocity vector of
the observers. The relations (\ref{P}) and (\ref{M}) are the basic relations
that are obtained in this paper. In our approach, the relations (\ref{P})
and (\ref{M}) replace the constitutive relations with the 3-vectors (\ref{pl}%
) and (\ref{mg}), which are equivalent to Minkowski's equations (\ref{de}).
The equations (\ref{pl}), (\ref{mg}) and (\ref{de}) are given below in Sec.
5.

In the special case when $v=u$, an observer frame comoving with medium, (\ref%
{P}) and (\ref{M}) yield
\begin{equation}
P=\varepsilon _{0}\chi _{E}E,\qquad M=(\chi _{B}/\mu _{0})B,  \label{pm}
\end{equation}%
which are the familiar linear constitutive laws for the case of an
electrically neutral, isotropic, non-dispersive, polarizable medium at rest.
Introducing the relative permittivity $\varepsilon _{r}$, $\varepsilon
_{r}=1+\chi _{E}$, and the relative permeability $\mu _{r}$, $\mu
_{r}=1/(1-\chi _{B})$, the constitutive laws (\ref{pm}) can be written as $%
P=\varepsilon _{0}(\varepsilon _{r}-1)E$, $M=(1/\mu _{0})(1-1/\mu _{r})B$.
In this case ($v=u$) the excitations $D$, $D=\varepsilon _{0}E+P$ and $H$, $%
H=(1/\mu _{0})B-M$ can be introduced, as in Sec. 2, and the rest frame
constitutive relations take the usual forms
\begin{equation}
D=\varepsilon E,\qquad H=(1/\mu )B,  \label{dh1}
\end{equation}%
where $\varepsilon =\varepsilon _{0}\varepsilon _{r}$ and $\mu =\mu _{0}\mu
_{r}$.

The last term in (\ref{P}) and (\ref{M}) describes the magnetoelectric
effects in a moving dielectric. According to the last term in (\ref{P}) a
moving dielectric becomes electrically polarized when placed in a magnetic
field, the Wilsons' experiment.$^{21}$ Similarly, according to the last term
in (\ref{M}) a moving dielectric becomes magnetized when it is placed in an
electric field, R\"{o}ntgen's experiment.$^{22}$\bigskip \bigskip

\noindent \textbf{4. Constitutive Relations for Moving Media in the Standard
Basis}\bigskip

\noindent The equations (\ref{cr1}) - (\ref{dh1}) are all coordinate-free
relations. In order to compare them with some usual formulations we have to
represent all 4D geometric quantities from them in the standard basis $%
\left\{ \gamma _{\mu }\right\} $. Then, in the $\left\{ \gamma _{\mu
}\right\} $ basis, Eq. (\ref{P}) becomes%
\begin{equation}
P^{\mu }\gamma _{\mu }=(\varepsilon _{0}\chi _{E}/c)[(1/c)(E^{\mu }v^{\nu
}-E^{\nu }v^{\mu })+\varepsilon ^{\mu \nu \alpha \beta }v_{\alpha }B_{\beta
}]u_{\nu }\gamma _{\mu },  \label{po}
\end{equation}%
whereas (\ref{M}) takes the form
\begin{equation}
M^{\mu }\gamma _{\mu }=\varepsilon _{0}\chi _{B}[(B^{\mu }v^{\nu }-B^{\nu
}v^{\mu })+(1/c)\varepsilon ^{\mu \nu \alpha \beta }E_{\alpha }v_{\beta
}]u_{\nu }\gamma _{\mu }.  \label{ma}
\end{equation}

We shall examine Eqs. (\ref{po}) and (\ref{ma}) taking that the laboratory
frame, the $S$ frame, is the $\gamma _{0}$-frame ($v=c\gamma _{0}$, $v^{\mu
}=(c,0,0,0)$, $E^{0}=B^{0}=0$) in which the material medium, the $S^{\prime
} $ frame, is moving with velocity $u$, $u^{\nu }=(\gamma _{u}c,\gamma
_{u}U^{1},\gamma _{u}U^{2},\gamma _{u}U^{3})$, $U^{k}$ are the components of
the 3-velocity $\mathbf{U}$ and $\beta _{u}=\left\vert \mathbf{U}\right\vert
/c$. Then, in the laboratory frame, Eq. (\ref{po}) becomes%
\begin{equation}
P=\varepsilon _{0}\chi _{E}\gamma _{u}[E^{i}(U^{i}/c)\gamma
_{0}+(E^{i}+\varepsilon ^{0ijk}U_{j}B_{k})\gamma _{i}].  \label{pc}
\end{equation}%
It can be seen from (\ref{pc}) that, e.g., if $E^{\mu }=(0,0,0,0)$, $B^{\mu
}=(0,0,0,-B^{3})$, $u^{\mu }=(\gamma _{u}c,\gamma _{u}U^{1},0,0)$, the
components of the polarization are
\begin{equation}
P^{\mu }=(0,0,P^{2}=\varepsilon _{0}\chi _{E}\gamma _{u}U^{1}B^{3},0);
\label{pwi}
\end{equation}%
these components correspond to the `translational' version of Wilsons'
experiment. As stated on page 13,$^{23}$ the magnetoelectric effect `in a
moving dielectric is of course anisotropic in the sense that the
polarization depends upon an applied field which is perpendicular to the
direction of motion and this polarization is then perpendicular to both the
applied field and the direction of motion.' Comparing $P^{2}$ from (\ref{pwi}%
) with $P_{y}$ from Eq. (1.14)$^{23}$ (see also the term $\chi
_{(em)}^{\alpha \beta }$ in Fig. 3.3 and Eq. (2.9)$^{23}$), we see that for
the above conditions (the electric field is absent) these two expressions
differ only in the $\gamma _{u}$ factor; $P^{2}$ contains $\gamma _{u}$
whereas $P_{y}$ contains $\gamma _{u}^{2}$.

However, it is worth mentioning that the complete $P_{y}$ (both, the
3-vectors of the electric and magnetic fields exist) from Eq. (1.14),$^{23}$
or the components of the 3-vector $\mathbf{P}$ from Eq. (2.9) and Fig. 3.3,$%
^{23}$ can be compared only with the spatial components of (\ref{pc}), but $%
P $ in (\ref{pc}) contains the temporal component as well. Again, the \emph{%
spatial components} of (\ref{pc}) and the complete $P_{y}$ differ only in
the $\gamma _{u}$ factor.

Similarly, from (\ref{ma}), we find
\begin{equation}
M=(\chi _{B}/\mu _{0})\gamma _{u}[B^{i}(U^{i}/c)\gamma
_{0}+(B^{i}-\varepsilon ^{0ijk}U_{j}E_{k}/c^{2})\gamma _{i}].  \label{ma1}
\end{equation}

In the case of low velocities of the medium, $\beta _{u}\ll 1$, the
equations (\ref{pc}) and (\ref{ma1}) are almost unchanged to first order of $%
\beta _{u}$; only it is taken that $\gamma _{u}\simeq 1$. Observe that in (%
\ref{pc}) and (\ref{ma1}) the terms with $\gamma _{0}$ are comparable to the
terms containing $U_{j}B_{k}\gamma _{i}$ and $(1/c^{2})U_{j}E_{k}\gamma _{i}$%
, respectively. Thus, to first order of $\beta _{u}$, the terms $P^{0}$ and $%
M^{0}$ \emph{cannot be neglected} relative to the complete $P^{i}$ and $%
M^{i} $, respectively.

The constitutive relations (\ref{pc}) and (\ref{ma1}) significantly differ
from all previous constitutive equations for such simple moving media. They
will be compared with the usual formulations with the 3-vectors.\bigskip
\bigskip

\noindent \textbf{5. Comparison with Minkowski's Constitutive Relations}

\textbf{and with Their Derivations\bigskip }

\noindent Let us examine some of the usual derivations of the constitutive
relations for moving media. As already mentioned, almost the entire physics
literature deals with the electromagnetic excitation tensor $\mathcal{H}$,
i.e. with the electric $D$ and magnetic $H$ excitations and the constitutive
relations refer to the connections between them and $F$, i.e. $E$ and $B$,
respectively.

Minkowski$^{1}$ was the first who derived the constitutive relations for the
3-vectors
\begin{eqnarray}
\mathbf{D}+(1/c^{2})\mathbf{U}\times \mathbf{H} &=&\varepsilon (\mathbf{E}+%
\mathbf{U}\times \mathbf{B})  \notag \\
\mathbf{B}-(1/c^{2})\mathbf{U}\times \mathbf{E} &=&\mu (\mathbf{H}-\mathbf{U}%
\times \mathbf{D}),  \label{de}
\end{eqnarray}%
Eqs. (C) and (D) in Sec. 8.$^{1}$ From that time, the equations (\ref{de})
are considered to be the fundamental constitutive equations for moving media
in the whole physics community. They are subsequently derived in different
ways and used in numerous papers and textbooks on the
electromagnetism.\bigskip \bigskip

\noindent \textbf{5.1\ Comparison with Minkowski's constitutive relations
that are}

\textbf{obtained by means of the 3-vectors}$^{24}$ \textbf{and}

\textbf{in the covariant approach}$^{25}$\textbf{\bigskip }

\noindent One way of the derivation of Eqs. (\ref{de}) is presented in Ref.
24. There, these equations, i.e. Eqs. (7),$^{24}$ are derived using
Minkowski's crucial hypothesis that the relations with the 3-vectors $%
\mathbf{D}=\varepsilon \mathbf{E}$, $\mathbf{H}=(1/\mu )\mathbf{B}$, which
correspond to (\ref{dh1}), retain their form in the moving frame $S^{\prime
} $ with the same $\varepsilon $ and $\mu $, i. e., $\mathbf{D}^{\prime
}=\varepsilon \mathbf{E}^{\prime }$, $\mathbf{H}^{\prime }=(1/\mu )\mathbf{B}%
^{\prime }$. Then, in Ref. 24, \emph{the UT} (\ref{JCB}) \emph{and the
similar UT for} $\mathbf{D}^{\prime }$ \emph{and} $\mathbf{H}^{\prime }$
(Eqs. (3)$^{24}$) \emph{are used to obtain Minkowski's constitutive equations%
} (\ref{de}). Furthermore, in Ref. 24, the constitutive relations, which
explicitly express $\mathbf{D}$ and $\mathbf{H}$ in terms of $\mathbf{E}$
and $\mathbf{B}$, are derived (Eqs. (10)$^{24}$) from (\ref{de}).
Introducing $\mathbf{D}=\varepsilon _{0}\mathbf{E}+\mathbf{P}$ and $\mathbf{H%
}=(1/\mu _{0})\mathbf{B}-\mathbf{M}$ into Eqs. (10)$^{24}$ we find the
constitutive equations which explicitly express $\mathbf{P}$ as a function
of $\mathbf{E}$ and $\mathbf{B}$
\begin{eqnarray}
\mathbf{P} &=&\gamma ^{2}\varepsilon _{0}\{\chi _{E}[\mathbf{E}-c^{-2}%
\mathbf{U}(\mathbf{U\cdot E})+\mathbf{U}\times \mathbf{B}]  \notag \\
&&+\chi _{B}[\mathbf{U\times (B}-c^{-2}\mathbf{U}\times \mathbf{E)]}\}
\label{pl}
\end{eqnarray}%
and also $\mathbf{M}$ in terms of $\mathbf{E}$ and $\mathbf{B}$
\begin{eqnarray}
\mathbf{M} &=&(\gamma ^{2}/\mu _{0})\{\chi _{B}[\mathbf{B}-c^{-2}\mathbf{U}(%
\mathbf{U\cdot B})-c^{-2}\mathbf{U}\times \mathbf{E}]  \notag \\
&&-c^{-2}\chi _{E}[\mathbf{U\times (E}+\mathbf{U}\times \mathbf{B)}]\}.
\label{mg}
\end{eqnarray}%
The equations (\ref{pl}) and (\ref{mg}) are equivalent to Minkowski's
equations (\ref{de}).

Introducing the relative permittivity $\varepsilon _{r}$ and the relative
permeability $\mu _{r}$ instead of the susceptibilities $\chi _{E}$ and $%
\chi _{B}$ the equations (\ref{pl}) and (\ref{mg}) become%
\begin{eqnarray}
\mathbf{P} &=&\varepsilon _{0}(\varepsilon _{r}-1)\mathbf{E}+\gamma
^{2}\varepsilon _{0}(\varepsilon _{r}-1/\mu _{r})\mathbf{U\times (B}-c^{-2}%
\mathbf{U}\times \mathbf{E),}  \notag \\
\mathbf{M} &=&\mu _{0}^{-1}(1-1/\mu _{r})\mathbf{B}-\gamma ^{2}\varepsilon
_{0}(\varepsilon _{r}-1/\mu _{r})\mathbf{U\times (E}+\mathbf{U}\times
\mathbf{B).}  \label{mp}
\end{eqnarray}%
and they are, in the same way as Eqs. (\ref{pl}) and (\ref{mg}), equivalent
to Minkowski's equations (\ref{de}).

Similarly, in Ref. 25, the relations (\ref{de}) (Eqs. (76.9)$^{25}$) are
obtained by the covariant generalization (only components, implicitly taken
in the standard basis) of the relations with the 3-vectors $\mathbf{D}%
=\varepsilon \mathbf{E}$, $\mathbf{H}=(1/\mu )\mathbf{B}$, which correspond
to our Eq. (\ref{dh1}). The covariant generalizations from Ref. 25 are $%
\mathcal{H}^{\lambda \mu }u_{\mu }=\varepsilon F^{\lambda \mu }u_{\mu }$
((76.7)$^{25}$) and $^{\ast }F^{\lambda \mu }u_{\mu }=\mu ^{\ast }\mathcal{H}%
^{\lambda \mu }u_{\mu }$ ((76.8)$^{25}$), where $^{\ast }F^{\lambda \mu }$
and $^{\ast }\mathcal{H}^{\lambda \mu }$ are the dual tensors, i.e., only
the components ($^{\ast }F^{\alpha \beta }=(1/2)\varepsilon ^{\alpha \beta
\gamma \delta }F_{\gamma \delta }$). Then, Minkowski's constitutive
equations (\ref{de}) with the 3-vectors are derived by \emph{Minkowski's
identifications,}$^{1}$\emph{\ of components of} $F^{\alpha \beta }$ \emph{%
with components of the 3-vectors} $\mathbf{E}$ \emph{and} $\mathbf{B}$ ($%
E^{i}=F^{i0}$, $B^{i}=(1/2c)\varepsilon ^{ijk0}F_{jk}$, see, e.g., Eq.
(11.137)$^{9}$) \emph{and by the similar identifications} \emph{for} $%
\mathcal{H}^{\alpha \beta }$ \emph{and the 3-vectors} $\mathbf{D}$ \emph{and}
$\mathbf{H}$ ($D^{i}=\mathcal{H}^{i0}$, $H^{i}=(c/2)\varepsilon ^{ijk0}%
\mathcal{H}_{jk}$).

However, as discussed, e.g. in Sec. 2,$^{11}$ the components are not the
whole physical quantity. The mentioned identifications of components are
synchronization dependent and the UT (\ref{JCB}) (and the similar ones for $%
\mathbf{D}^{\prime }$ and $\mathbf{H}^{\prime }$) significantly differ from
the LT (\ref{T1}).

On the other hand, our equations (\ref{cr1}) and (\ref{cr2}) deal with
coordinate-free, 4D geometric quantities. Instead of Minkowski's
identifications of components, and the same ones in Ref. 25, the
mathematically correct decompositions of $F$ (\ref{E2}) and $\mathcal{M}$ (%
\ref{M1}) are used for the derivation of the constitutive equations (\ref{P}%
) and (\ref{M}) (and, in the standard basis, (\ref{po}) and (\ref{ma})) from
(\ref{cr1}) and (\ref{cr2}), respectively.

As mentioned above, the equations (\ref{pl}) and (\ref{mg}), or (\ref{mp}),
which are equivalent to (\ref{de}), can be compared with Eqs. (\ref{pc}) and
(\ref{ma1}). There are important differences between them.

1) The equations (\ref{pc}) and (\ref{ma1}) are with correctly defined 4D
quantities that properly transform under the LT, whereas it is not the case
with Eqs. (\ref{pl}) and (\ref{mg}), or (\ref{mp}), with the 3-vectors that
transform according to the UT.

2) Furthermore, (\ref{pc}) and (\ref{ma1}) contain the term with $\gamma
_{0} $, which cannot exist in the approach with the 3-vectors, i.e. in (\ref%
{pl}) and (\ref{mg}), or (\ref{mp}).

3) Also, in (\ref{pc}), the polarization vector $P$ contains only the
electric susceptibility $\chi _{E}$ and not $\chi _{B}$, whereas, in (\ref%
{pl}), the polarization 3-vector $\mathbf{P}$ contains both susceptibilities
$\chi _{E}$ and $\chi _{B}$, (or, in (\ref{mp}), both, $\varepsilon _{r}$
and $\mu _{r}$). Similarly, in (\ref{ma1}), there is only $\chi _{B}$ and
not $\chi _{E}$, whereas, in (\ref{mg}), there are both susceptibilities $%
\chi _{B}$ and $\chi _{E}$, (or, in (\ref{mp}), both, $\mu _{r}$ and $%
\varepsilon _{r}$ ). It can be seen from the derivation of (\ref{pl}) and (%
\ref{mg}), or (\ref{mp}), that both susceptibilities appear in the
expressions for $\mathbf{P}$ (\ref{pl}) and $\mathbf{M}$ (\ref{mg}) because
the UT (\ref{JCB}) and the similar ones for $\mathbf{D}^{\prime }$ and $%
\mathbf{H}^{\prime }$ (Eqs. (3)$^{24}$), or, equivalently, the UT (\ref{ps}%
), are used and in them, e.g., in (\ref{JCB}), $\mathbf{E}^{\prime }$ \emph{%
is expressed by the mixture of }$\mathbf{E}$ \emph{and} $\mathbf{B}$,\ and
similarly, in (\ref{ps}), $\mathbf{P}^{\prime }$ \emph{is expressed by the
mixture of }$\mathbf{P}$ \emph{and} $\mathbf{M}$.\bigskip \bigskip

\noindent \textbf{5.2. Comparison of the low velocity limits\bigskip }

\noindent These differences remain in the low velocity limit, $\beta _{u}\ll
1$, as well. That limit is already examined for vectors $P$ and $M$ from (%
\ref{pc}) and (\ref{ma1}), respectively. Only $\gamma _{u}\simeq 1$ is taken
in (\ref{pc}) and (\ref{ma1}).

In Ref. 24, the low velocity limit, i.e., the quasi-static approximation, is
obtained in the same way by putting $\gamma _{u}\simeq 1$ in Eqs. (10) for $%
\mathbf{D}$ and $\mathbf{H}$ expressed in terms of $\mathbf{E}$ and $\mathbf{%
B}$, which led to Eqs. (11).$^{24}$ They, Eqs. (11),$^{24}$ are commonly
used in literature. In the formulation with the 3-vectors $\mathbf{P}$ and $%
\mathbf{M}$ the equations (\ref{mp}) are derived from Eqs. (10).$^{24}$
Therefore, in that approximation, the usual 3-vectors $\mathbf{P}$ and $%
\mathbf{M}$ are given by (\ref{mp}) but with $\gamma _{u}\simeq 1$. This
means that again both $\varepsilon _{r}$ and $\mu _{r}$, i.e., both
susceptibilities $\chi _{E}$ and $\chi _{B}$, appear in the quasi-static
approximation in the expressions for $\mathbf{P}$ and $\mathbf{M}$. In Ref.
24, it is argued that such a procedure with $\gamma _{u}\simeq 1$ does not
give a correct quasi-static approximation, because the obtained equations,
(11),$^{24}$ `do not obey the group additivity property.' Hence, according
to Ref. 24, the equations (11)$^{24}$ have to be replaced by two
well-defined Galilean limits of Minkowski's constitutive equations ((\ref{de}%
) here), the magnetic and electric limits, i.e. with two sets of
low-velocity formulae. These two sets are Eqs. (12) and (13) in Ref.24 and
it is stated there: `Both sets (12) and (13) do form a group.'

However, the UT (\ref{JCB}) and (\ref{ps}) and the similar UT for $\mathbf{D}%
^{\prime }$ and $\mathbf{H}^{\prime }$ (Eqs. (3)$^{24}$) are not the LT and
therefore constitutive equations with the 3-vectors, (\ref{de}), or Eqs.
(10),$^{24}$, and also (\ref{pl}) and (\ref{mg}), or (\ref{mp}), here, are
not the \emph{relativistic} constitutive equations and also Eqs. (12) and
(13)$^{24}$ are not a quasi-static approximation of the relativistically
correct constitutive relations. Thus, as stated above, all differences from
Sec. 5.1 remain in the low velocity limit as well.

In the special case of moving ($\beta _{u}\ll 1$) nonmagnetic ($\mu _{r}=1$,
i.e. $\chi _{B}=0$) media the 3D polarization $\mathbf{P}$ from (\ref{mp}),
to first order of $\beta _{u}$, becomes

\begin{equation}
\mathbf{P=}\varepsilon _{0}\chi _{E}(\mathbf{E}+\mathbf{U}\times \mathbf{B}).
\label{pc1}
\end{equation}%
The equation (\ref{pc1}) is Eq. (9-19) (2),$^{19}$ or the last equation in
Problem 6.8.$^{20}$

But, even for $\beta _{u}\ll 1$, $P$ from (\ref{pc}) will have the same form
for both cases $\chi _{B}\neq 0$ and $\chi _{B}=0$, because it does not
depend on $\chi _{B}$ ($P\simeq \varepsilon _{0}\chi
_{E}[E^{i}(U^{i}/c)\gamma _{0}+(E^{i}+\varepsilon ^{0ijk}U_{j}B_{k})\gamma
_{i}]$). As already stated, to first order of $\beta _{u}$, $P^{0}$ cannot
be neglected relative to $P^{i}$, which means that we have not an equation
that would correspond to (\ref{pc1}).\bigskip \bigskip

\noindent \textbf{5.3. Comparison with constitutive equations that are
obtained}

\textbf{by means of exterior forms}$^{2}$\textbf{\bigskip }

\noindent It is worth noting that Minkowski's constitutive equations (\ref%
{de}) are also obtained in Ref. 2 using exterior calculus, i.e., with the
use of abstract 4D geometric quantities, Eqs. (E.4.28) and (E.4.29).$^{2}$
At that place (p. 353) Hehl and Obukhov stated: `Originally, the
constitutive relations (E.4.28), (E.4.29) were derived by Minkowski with the
help of Lorentz transformations for the case of a flat space time and a
uniformly moving medium. We stress, however, that the Lorentz group never
entered the scene in our derivation above.' It can be concluded from these
statements that the authors$^{2}$ believe, as all others, that the UT (\ref%
{JCB}) are the relativistically correct LT. However, as discussed in Sec. 2,
the LT are given by Eq. (\ref{T1}) and not by (\ref{JCB}). Furthermore, in
Ref. 2, the expressions for the polarization $P$ and the magnetization $M$,
the equations (E.4.30) and (E.4.31), respectively, are derived from the
constitutive relations (E.4.28), (E.4.29). The expressions for $P$ and $M,$
(E.4.30) and (E.4.31), respectively, from Ref. 2, are very similar to the
equations with the 3-vectors, (\ref{pl}) and (\ref{mg}), respectively, which
are obtained from (\ref{de}). Indeed, all quantities in the constitutive
relations (E.4.28), (E.4.29) and in the expressions for $P$ and $M$,
(E.4.30) and (E.4.31), respectively, from Ref. 2, are in the 3D subspace,
i.e., like the usual 3-vectors. (Observe that $v$ that enters into these
relations is the 3-velocity 1-form, (E.4.27), or the 3-velocity vector.
Also, the Hodge star operators in these expressions are defined by the
3-space metrics of the laboratory and material foliations.) This means that
these results$^{2}$ also significantly differ from the constitutive
relations (\ref{P}) and (\ref{M}) in which all quantities are the 4D
geometric quantities and there is no the space-time split. Let us describe
in more detail the calculation from Ref. 2, which led to the mentioned
results.$^{2}$

In Ref. 2, Minkowski's constitutive relations (E.4.28), (E.4.29) and the
expressions for $P$ and $M$, (E.4.30) and (E.4.31), respectively, are
derived using (1+3) - splitting of spacetime both in the laboratory frame
and in the frame of moving macroscopic matter, the material frame. The
starting relations in that derivation are the decompositions ((1+3) -
splitting) of the electromagnetic excitation tensor $\mathcal{H}$ and of $F$
in the laboratory frame (unprimed quantities), ($\mathcal{H=}-H\wedge dt+D$,
$F=E\wedge dt+B$) Eq. (E.4.14), and in the material frame (primed
quantities), ($\mathcal{H=}-H^{\prime }\wedge dt^{\prime }+D^{\prime }$, $%
F=E^{\prime }\wedge dt^{\prime }+B^{\prime }$) Eq. (E.4.15). (In both
equations our notations is used.) There is an assumption in connection with
the decompositions (E.4.14) and (E.4.15) in Ref. 2. It is: `Clearly, we
preserve the same symbols $\mathcal{H}$ (our notation) and $F$ on the
left-hand sides of (E.4.14) and (E.4.15) because these are just the same
physical objects. In contrast, the right-hand sides are of course different,
hence we use primes.'

From the mathematical point of view there are no reasons for such
argumentations. If $H$, $dt$, $D$, $E$, $B$ are transformed then $\mathcal{H}
$ and $F$ have to be transformed as well, as can be concluded going to a
uniformly moving medium, i.e., when the LT are applicable. Let us explain
these assertions in more detail. \emph{There is no mathematical procedure by
which in one equation some parts of it are transformed and the other parts
are not. It is not possible that there is one law for the transformations of
some 4D geometric quantities and another law for the transformations of the
other quantities. }As mentioned, e.g., in Ref. 11, \textquotedblleft \emph{%
any multivector} $M$ \emph{transforms by the active LT in the same way}, Eq.
(11),$^{11}$ $M\rightarrow M^{\prime }=RM\widetilde{R}$, where the Lorentz
boost $R$ is given by Eq. (10).$^{11}$ Hence, if $R$ is applied to (\ref{E2}%
), $R(F=E\wedge v/c+(IcB)\cdot v/c)\widetilde{R}$, then $F$ \emph{is also
transformed}, together with $E$, $B$ and $v$. Thus, the assertion that $F$
is unchanged under the active LT is without any mathematical justification.

In addition, let us compare the relations (E.4.14) and (E.4.15) with the
consideration from Ref. 26. In Sec. 4 `Premetric electrodynamics in exterior
calculus' in Ref. 26, it is argued that the decomposition of $\mathcal{H}$,
Eq. (31),$^{26}$ if put in matrix form, yields the identification (6)$_{1}$
in Ref. 26. Furthermore, the field strength $F$ is decomposed in the same
way as $\mathcal{H}$, i.e. `into two pieces: one along the 1-dimensional
time $t$ and another one embedded in 3-dimensional space ($a,b=1,2,3.$).'
Thus, in the notation,$^{26}$ $F=E\wedge dt+B=E_{a}dx^{a}\wedge
dt+(1/2)B_{ab}dx^{a}\wedge dx^{b}$, Eq. (34).$^{26}$ If put in matrix form,
it yields the usual identification (6)$_{2}$ in Ref. 26, i.e., in our
notation, $E^{i}=F^{i0}$, $B^{i}=(1/2c)\varepsilon ^{ijk0}F_{jk}$. The
decompositions (31) and (34)$^{26}$ are the same as the decompositions
(E.4.14).$^{2}$ As already stated, in the material frame, the same
decompositions of $\mathcal{H}$ and $F$ ($F=E^{\prime }\wedge dt^{\prime
}+B^{\prime }$) are performed, (E.4.15),$^{2}$ and they yield the usual
identifications (6)$_{1}$ and (6)$_{2}$ in Ref. 26, but now with primes, $%
E^{\prime i}=F^{\prime i0}$, $B^{\prime i}=(1/2c)\varepsilon
^{ijk0}F_{jk}^{\prime }$. This discussion reveals that the (1+3) - splitting
of spacetime, i.e., \emph{the usual identifications} (6)$_{1}$, (6)$_{2}$
\emph{in both frames}, are the real cause that, although the authors$^{2}$
worked with exterior forms in the 4D spacetime, they obtained the same
results for Minkowski's constitutive relations and for $D$ and $H$ ((E.4.25)
and (E.4.26)$^{2}$), i.e., for $P$ and $M$, as those that were found by
means of the 3-vectors and their UT, e.g., in Ref. 24.

However, as already discussed, e.g., in Refs. 27 and 11, and also in Ref.
12, \emph{the (1+3) - splitting of spacetime and the usual identifications
of components of, e.g.,} $F^{\alpha \beta }$ \emph{(implicitly taken in the
standard basis) with components of the 3D} $\mathbf{E}$ \emph{and} $\mathbf{B%
}$ \emph{in both frames are synchronization dependent and they are
meaningless in the `r' synchronization.} We note that \emph{different
synchronizations are nothing else than different conventions and physics
must not depend on conventions. }(In Ref. 26, Hehl (Eq. (6)$_{2}$) quotes
Minkowski's identifications from Sec. 3$^{1}$ and states: `We stress that
the identifications in (6) are premetric, they are valid on any
(well-behaved) differential manifold that can be split locally into time and
space.' This means that Hehl considers that 1+3 decomposition, i.e., the
space-time split, is also - premetric. But, if a nonstandard basis, the $%
\{r_{\mu }\}$ basis with the `r' synchronization is used, i.e., when the
appropriate metric is used, then it is not possible to make the usual
identifications, that is, 1+3 decomposition, i.e., the space-time split.
That metric is discussed in, e.g., the text below Eq. (14)$^{12}$: `.. the
components $g_{\mu \nu ,r}$ of the metric tensor $g_{ab}$ are $g_{ii,r}=0$,
and all other components are $=1$.'\ Hence, both, the usual identifications
(Eqs. (6)$^{26}$), and 1+3 decomposition, i.e., the space-time split, are
meaningful \emph{only} when the Minkowski metric, e.g., $diag(1,-1,-1,-1)$,
is used. Thus, these identifications depend on the chosen metric and
therefore they are not premetric.

(For the $\{r_{\mu }\}$ basis see, for example, Refs. 27, 12.) In the $%
\{r_{\mu }\}$ basis, it holds that $F_{r}^{10}=E^{1}+cB^{3}-cB^{2}$. The
relation for $F_{r}^{10}$ shows that \emph{the components have no definite
physical meaning }since they are dependent on the chosen synchronization.
Only in the $\{\gamma _{\mu }\}$ basis does it hold that $E^{i}=F^{i0}$, $%
P^{i}=\mathcal{M}^{i0}$, etc. According to that, the usual 3-vectors $%
\mathbf{E}$, $\mathbf{B}$, $\mathbf{D}$, $\mathbf{H}$, $\mathbf{P}$, $%
\mathbf{M}$, etc., where, e.g., $\mathbf{P=}\mathcal{M}^{10}\mathbf{i}+%
\mathcal{M}^{20}\mathbf{j}+\mathcal{M}^{30}\mathbf{k}$ ($\mathbf{i}$, $%
\mathbf{j}$, $\mathbf{k}$ are the unit 3-vectors) \emph{have no definite
physical meaning, }since the components $\mathcal{M}^{i0}$ are dependent on
the chosen synchronization.)

Comparing, for example, the decompositions of $F$ from Ref. 2 and 26 (the
above quoted Eq. (34)$^{26}$) with our decomposition $F=E\wedge
v/c+(IcB)\cdot v/c$, Eq. (\ref{E2}), i.e., in some arbitrary basis $\left\{
e_{\mu }\right\} $ (it can be the standard basis $\left\{ \gamma _{\mu
}\right\} $, or the $\left\{ r_{\mu }\right\} $ basis with the `r'
synchronization, etc.) $F=(\delta _{\quad \mu \nu }^{\alpha \beta
}E_{e}^{\mu }v_{e}^{\nu }+c\varepsilon ^{\alpha \beta \mu \nu }v_{e,\mu
}B_{e,\nu })e_{\alpha }\wedge e_{\beta }$ ($\alpha ,\beta ,\mu ,\nu =0,1,2,3$
and $E_{e}^{\mu }$, $B_{e}^{\mu }$, $v_{e}^{\mu }$ are the components in the
$\left\{ e_{\mu }\right\} $ basis), we see that there are important
differences between them. All quantities, including $v$, in our relations
are the 4D quantities and there is no (1+3) - splitting of spacetime, which
means that they are more general than the decompositions from Ref. 2 and 26.
If, for example, the magnetization 1-form $M$, Eq. (E.4.31),$^{2}$ would be
written in the $\left\{ \gamma _{\mu }\right\} $ basis then it would contain
in the material frame \emph{and in the laboratory frame} only spatial
components and not the time component in contrast to the relation (\ref{ma1}%
) (and (\ref{pc})). Observe also that relations (\ref{P}) and (\ref{M}), or
in the $\left\{ \gamma _{\mu }\right\} $ basis (\ref{po}) and (\ref{ma}),
contain both the velocity of the observer $v$ and the velocity of the
considered medium $u$, whereas there is only one velocity, the velocity of
the considered medium, in the equations (E.4.25)-(E.4.31).$^{2}$

Furthermore, the obvious similarity of the results with exterior forms$^{2}$
and the usual results with the 3-vectors and their UT, e.g. from Ref. 18 (or
24), can be nicely seen comparing Sec. E.4.5 under the title `The
experiments of R\"{o}ntgen and Wilson$\And $Wilson' in Ref. 2 and Secs. 5.2
and 5.3 in Ref. 18. It is visible from Eq. (E.4.49) and Fig. E.4.2$^{2}$
that in the matter-free region, region (2), there is only an electric field $%
E_{(2)}$, whereas there are \emph{both} the electric field $E_{(1)}$ and
\emph{the induced magnetic field} $B_{(1)}$ in the slab moving with \emph{%
constant} velocity. Completely the same result for that case is obtained in
the third equation in (5.18) in Sec. 5.2,$^{18}$ where it is exclusively
dealt with the 3-vectors and their UT. The same conclusion holds for the
comparison of the treatments of the Wilsons' experiment, Sec. E.4.5$^{2}$
and Sec. 5.3.$^{18}$

From the above discussion it is visible that either by using the 3-vectors
and their UT, or by using the modern mathematical language, the exterior
forms, always, in all previous treatments, it is obtained that, e.g. the
electric field in one frame is `seen' as slightly changed electric field
\emph{and} \emph{the induced magnetic field} in the frame relatively moving
with \emph{constant} velocity. The same holds for the polarization and the
magnetization and also for the electric and magnetic excitations.\bigskip
\bigskip

\noindent \textbf{6. The Physical Explanation of the Magnetoelectric Effect}

\textbf{in Moving Media\bigskip }

\noindent The consideration in the previous sections shows that there are
important differences in the form of Minkowski's constitutive relations (\ref%
{de})\textbf{,} or equivalently the constitutive relations for $\mathbf{P}$ (%
\ref{pl}) and $\mathbf{M}$ (\ref{mg}), i.e., (\ref{mp}), and the
constitutive relations for vectors $P$ (\ref{P}) and $M$ (\ref{M}). However,
there is also a significant difference in physical interpretation of these
constitutive relations. Particularly, this refers to the interpretation of
the magnetoelectric effect for moving media.

In previous approaches it is considered that the term which describes how
the magnetic field influences the polarization ($\gamma ^{2}\varepsilon
_{0}(\varepsilon _{r}-1/\mu _{r})\mathbf{U\times B}$ in Eq. (\ref{mp})) and
the term which describes how the electric field influences the magnetization
($-\gamma ^{2}\varepsilon _{0}(\varepsilon _{r}-1/\mu _{r})\mathbf{U\times E}
$ in Eq. (\ref{mp})) are determined by the UT (\ref{JCB}) for $\mathbf{E}$
and $\mathbf{B}$. Namely, according to the UT (\ref{JCB}), in the rest frame
of a moving medium the magnetic field 3-vector from the laboratory frame is
`seen' as a slightly changed magnetic field 3-vector and an \emph{induced
electric field 3-vector}. That induced electric field 3-vector interacts
with the electric dipole moments 3-vectors by the interaction term $\mathbf{E%
}\cdot \mathbf{d}$ giving the polarization $\mathbf{P}$ as a 3-vector.
Similarly, in the rest frame of a moving medium the electric field from the
laboratory frame is `seen' as a slightly changed electric field and an \emph{%
induced magnetic field}. That induced magnetic field interacts with the
magnetic dipole moments by the interaction term $\mathbf{B}\cdot \mathbf{m}$
giving the magnetization $\mathbf{M}$ (all are 3-vectors). In my opinion, it
is very strange that, e.g., the magnetic field from a permanent magnet is
`seen' as an \emph{induced electric field }in a relatively moving frame.
What is the physical reason which causes that one field becomes another
field in a relatively moving frame? How can it be that a mere uniform moving
transforms one field to another one? It cannot be that the transformations
describe the physics, but, on the contrary, the physics has to describe how
to get the correct transformations of the fields.

If the electric and magnetic fields are interpreted as vectors defined on
the 4D spacetime then it is very natural to have that a magnetic field
vector remains the magnetic field vector in a relatively moving frame. There
is no physical cause that can change one field to another one in a
relatively moving frame. According to the LT (\ref{T1}), an electric field
vector transforms again to the electric field vector and similarly for a
magnetic field vector. Hence, the explanation for the magnetoelectric effect
in moving media is completely different in our approach.

Instead of dealing with the electric and magnetic dipole moments 3-vectors $%
\mathbf{d}$ and $\mathbf{m}$ we deal with the dipole moment bivector $D$, as
a primary quantity for dipole moments, which does have a definite physical
reality. The decomposition of $D$ into the electric and magnetic dipole
moments vectors $d$ and $m$, respectively, and the unit time-like vector $%
u/c $, and the expressions for $d$ and $m$, which are obtained from $D$ and
determined relative to $u$ are%
\begin{eqnarray}
D &=&d\wedge u/c+(mI)\cdot u/c^{2},  \notag \\
d &=&D\cdot u/c,\quad m=cI(D\wedge u/c);  \label{di}
\end{eqnarray}%
compare with (\ref{M1}) and (\ref{M2}). In this case the vector $u$ is the
bulk velocity vector of the medium as in (\ref{M1}) and (\ref{M2}). The
interaction term in the Lagrangian for the interaction between the
electromagnetic field $F$ and the dipole moment bivector $D$ can be written
as a sum of two terms%
\begin{eqnarray}
L_{int} &=&F\cdot D=(1/c^{2})[-(E\cdot d+B\cdot m)(v\cdot u)+(E\cdot
u)(v\cdot d)  \label{i} \\
&&+(B\cdot u)(v\cdot m)]+(1/c^{3})[(E\wedge m-c^{2}B\wedge d)\wedge v\wedge
u]I.  \notag
\end{eqnarray}%
In the tensor formulation, the relations (\ref{di}) and (\ref{i}) are given
by Eqs. (2) and (3), respectively, in Ref. 12 (they are first reported in
Ref. 28). Observe that every term on the r.h.s. of (\ref{i}) contains both
velocities $u$ and $v$. As seen from the last two terms they contain the
direct interaction of $E$ with $m$, and $B$ with $d$. \emph{These terms give
the physical explanation for the existence of the magnetoelectric effect in
moving media.} Moreover, there is no need for any transformation. We only
need to represent $E$, $d$, $B$, $m$, $u$ and $v$ from (\ref{i}) in the
standard basis and then to choose the laboratory frame as our $\gamma _{0}$%
-frame. It can be seen from the discussion of Eq. (25)$^{12}$ that in the
laboratory frame, as the $\gamma _{0}$-frame, and in the low velocity limit,
we can neglect the contributions to $L_{int}$ from the terms with $d^{0}$
and $m^{0}$; they are $u^{2}/c^{2}$ of the usual terms $E\cdot d$ or $B\cdot
m$. Then, what remains from (\ref{i}) is
\begin{equation}
L_{int}=-((E_{i}d^{i})+(B_{i}m^{i}))-(1/c^{2})\varepsilon
^{0ijk}(E_{i}m_{k}-c^{2}B_{i}d_{k})u_{j}.  \label{1}
\end{equation}%
This is, to order $0(u^{2}/c^{2})$, relativistically correct expression with
vectors for $L_{int}$. The last two terms that contain the direct
interactions between $E$ and $m$ and between $B$ and $d$ are not taken into
account in any of the previous investigations of the magnetoelectric effect
in moving media. They are $u/c$ of the usual terms with the direct
interaction of $E$ with $d$ and $B$ with $m$. (The expression (\ref{1}) is
first reported in Ref. 29 in connection with the EDM searches.)\bigskip
\bigskip

\noindent \textbf{7. A Brief Discussion of the General Constitutive
Relations\bigskip }

\noindent In this paper, the consideration is restricted to the constitutive
relations and the magnetoelectric effect in moving media with homogeneous
and isotropic electric and magnetic properties. For them, the rest frame
constitutive relations are given by Eqs. (\ref{dh1}). The general
constitutive relations which cover dielectric, magnetic and magnetoelectric
behavior are considered, e.g., in Ref. 26 and in more detail in Refs. 30, 2,
31 and 23. Mainly, e.g., Refs. 2, 26, 30, 31, these more general
constitutive relations link the electromagnetic excitation $\mathcal{H}$ and
$F$, $\mathcal{H}_{\alpha \beta }=\kappa _{\alpha \beta }{}^{{}\gamma \delta
}F_{\gamma \delta }$ (our notation), Eq. (17),$^{30}$ where `$\kappa
_{\alpha \beta }{}^{{}\gamma \delta }(x)$ is the twisted constitutive tensor
of type $\left[
\begin{array}{c}
2 \\
2%
\end{array}%
\right] $' or, in another form, e.g., $^{\ast }\mathcal{H}^{\alpha \beta
}=\chi ^{\alpha \beta \gamma \delta }F_{\gamma \delta }$ (our notation), Eq.
(21),$^{26}$ where $^{\ast }\mathcal{H}^{\alpha \beta }=(1/2)\varepsilon
^{\alpha \beta \gamma \delta }\mathcal{H}_{\gamma \delta }$, $\chi ^{\alpha
\beta \gamma \delta }$ is `a \textit{constitutive tensor density} of rank 4
and weight +1, with the dimension [$\chi $]=1/resistance.' In Refs. 23, the
general constitutive relations are established by expressing $\mathcal{M}$
as a linear function of the electromagnetic field $F$, $\mu _{0}c\mathcal{M}%
_{\alpha \beta }=(1/2)\xi _{\alpha \beta }^{\gamma \delta }F_{\gamma \delta
} $ (our notation), Eq. (2.45),$^{23}$ where `$\xi _{\alpha \beta }^{\gamma
\delta }$ is the \emph{general susceptibility tensor}, which is
dimensionless because of the choice of the constant $\mu _{0}c$.' However,
in all these treatments, all quantities in the equations are not geometric
quantities but components in the standard basis, which means that the
components of $F$, $\mathcal{H}$, $\mathcal{M}$ are obtained by the usual
identifications, i.e., they are considered to be the components of the
3-vectors $\mathbf{E}$ and $\mathbf{B}$, $\mathbf{D}$ and $\mathbf{H}$, $%
\mathbf{P}$ and $\mathbf{M}$, respectively. This can be nicely seen from
Eqs. (2.47) and (2.48)$^{23}$ and from their comparison with Eq. (2.9).$%
^{23} $ Hence, these relations are not so general relations and they are not
premetric, as stated, e.g., in Ref. 26, because, as already mentioned in
Sec. 5.3, the usual identifications and the space-time split are
meaningless, e.g., in the $\{r_{\mu }\}$ basis with the `r'\ synchronization.

In contrast with the usual covariant approach with coordinate-dependent
quantities, all relations (\ref{E2}) - (\ref{h1}), (\ref{T1}), (\ref{cr1}) -
(\ref{dh1}) and (\ref{di}) - (\ref{i}) are coordinate-free relations, which
are written in terms of the abstract 4D geometric quantities. \bigskip
\bigskip

\noindent \textbf{8. Conclusions}\bigskip

\noindent The constitutive relations for $P$ (\ref{P}) and $M$ (\ref{M})
contain both velocity vectors $u$ and $v$ and thus they differ from all
previous expressions. They are formulated in terms of coordinate-free
quantities that correctly transform under the LT (\ref{T1}), whereas, as
explained in Secs. 5 - 5.2, it is not the case with Minkowski's constitutive
relations (\ref{pl}) and (\ref{mg}), or (\ref{mp}), with the 3-vectors that
transform according to the UT (\ref{JCB}) and (\ref{ps}), which are not the
LT. As discussed in Sec. 5.3, Minkowski's constitutive relations for $P$ and
$M$, (E.4.30) and (E.4.31), respectively, from Ref. 2, are obtained by the
(1+3) - splitting of spacetime and consequently they are equivalent to the
usual expressions with the 3-vectors (\ref{pl}) and (\ref{mg}). Hence, these
relations,$^{2}$ which are obtained using exterior forms, also differ from
our results (\ref{P}) and (\ref{M}). The differences that are quoted at the
end of Sec. 5.1, points 1) - 3), could be used for the experimental
examination and comparison of the results presented here and the
constitutive relations which are obtained in the usual formulations either
with the 3-vectors or with exterior forms. Furthermore, a completely new
physical explanation of the magnetoelectric effect in moving media that is
presented in Sec. 6, Eq. (\ref{i}), or Eq. (\ref{1}), offers the possibility
for the experimental investigations of the magnetoelectric effect from the
relativistically correct point of view. Regarding the importance of the
magnetoelectric effect the results obtained in this paper could be enough
important in different physical applications.\bigskip \bigskip

\noindent \textbf{References\bigskip }

\noindent \lbrack 1] H. Minkowski, \textit{Nachr. Ges. Wiss. G\"{o}ttingen},
53 (1908);

Reprinted in: \textit{Math. Ann.} \textbf{68,} 472 (1910);

English translation in: M. N. Saha and S. N. Bose \textit{The Principle }

\textit{of Relativity: Original Papers by A. Einstein and H. Minkowski}

(Calcutta University Press, Calcutta, 1920).

\noindent \lbrack 2] F. W. Hehl and Yu. N. Obukhov, \textit{Foundations of
Classical }

\textit{Electrodynamics: Charge, flux, and metric} (Birkh\"{a}user, Boston,
2003).

\noindent \lbrack 3] T. Ivezi\'{c}, \textit{arXiv: 1101.3292}.

\noindent \lbrack 4] T. Ivezi\'{c}, \textit{Found. Phys. Lett.} \textbf{18},
401 (2005).

\noindent \lbrack 5] D. Hestenes, \textit{Space-Time Algebra (}Gordon \&
Breach, New York, 1966);

D. Hestenes and G. Sobczyk, \textit{Clifford Algebra to }

\textit{Geometric Calculus }(Reidel, Dordrecht, 1984);

C. Doran and A. Lasenby, \textit{Geometric algebra for physicists}

(Cambridge University Press, Cambridge, 2003 ).

\noindent \lbrack 6] A. Einstein, \textit{Annalen der Physik} \textbf{17,}
891 (1905); Translated by W. Perrett

and G. B. Jeffery in: \textit{The Principle of Relativity} (Dover, New York,
1952).

\noindent \lbrack 7] C. M\o ller, \textit{The Theory of Relativity} 2nd ed.
(Clarendon Press,

Oxford, 1972); P. Hillion, \textit{Phys. Rev. E} \textbf{48,} 3060 (1993);

T. Dereli, J. Gratus and R. W. Tucker, \textit{Phys. Lett. A} \textbf{361,}
190 (2007).

\noindent \lbrack 8] Shin-itiro Goto, R.W. Tucker and T.J. Walton, \textit{%
arXiv: 1003.1637}.

\noindent \lbrack 9] J.D. Jackson, \textit{Classical Electrodynamics} 3rd
ed. (Wiley, New York, 1998).

\noindent \lbrack 10] Z. Oziewicz, \textit{Rev. Bull. Calcutta Math. Soc}.%
\textit{\ }\textbf{16}, 49 (2008);

Z. Oziewicz and C.K. Whitney, \textit{Proc. Nat. Phil. Alliance}

\textit{(NPA)} \textbf{5}, 183 (2008), (also at
http://www.worldnpa.org/php/);

Z. Oziewicz, \textit{J. Phys.: Conf. Ser.} \textbf{330}, 012012 (2011).

\noindent \lbrack 11] T. Ivezi\'{c}, \textit{Phys. Scr.} \textbf{82}, 055007
(2010).

\noindent \lbrack 12] T. Ivezi\'{c}, \textit{Phys. Scr.} \textbf{81}, 025001
(2010).

\noindent \lbrack 13] T. Ivezi\'{c}, \textit{Found. Phys.} \textbf{33}, 1339
(2003).

\noindent \lbrack 14] T. Ivezi\'{c}, \textit{Found. Phys. Lett.} \textbf{18}%
, 301 (2005).

\noindent \lbrack 15] T. Ivezi\'{c}, \textit{Found. Phys.} \textbf{35}, 1585
(2005).

\noindent \lbrack 16] T. Ivezi\'{c}, \textit{Fizika A\ }\textbf{17}, 1
(2008).

\noindent \lbrack 17] T. Ivezi\'{c}, \textit{arXiv:\ 0809.5277}.

\noindent \lbrack 18] J. Van Bladel, \textit{Relativity and Engineering}
(Springer-Verlag, Berlin, 1984).

\noindent \lbrack 19] W.K.H. Panofsky and M. Phillips, \textit{Classical
electricity and magnetism} 2nd

ed. (Addison-Wesley, Reading, 1962).

\noindent \lbrack 20] W.G.W. Rosser, \textit{Classical Electromagnetism via
Relativity }(Plenum,

New York, 1968).

\noindent \lbrack 21] M. Wilson and H.A. Wilson, \textit{Proc. Roy. Soc.
(London)} \textit{A}\textbf{89}, 99 (1913).

\noindent \lbrack 22] W.C. R\"{o}ntgen, \textit{Ann. Phys. (Leipzig)}
\textbf{35}, 264 (1888).

\noindent \lbrack 23] T.H. O'Dell, \textit{The Electrodynamics of
Magneto-Electric Media}

(North-Holland, Amsterdam, 1970).

\noindent \lbrack 24] G. Rousseaux, \textit{Europhys. Lett.} \textbf{84},
20002 (2008).

\noindent \lbrack 25] L. Landau and E. Lifshitz, \textit{Electrodynamics of
Continuous Media},

2nd ed. (Pergamon, New York, 1984).

\noindent \lbrack 26] F.W. Hehl, \textit{Annalen der Physik} \textbf{17},
691 (2008).

\noindent \lbrack 27] T. Ivezi\'{c}, \textit{Found. Phys.} \textbf{31}, 1139
(2001).

\noindent \lbrack 28] T. Ivezi\'{c}, \textit{Phys. Rev. Lett.} \textbf{98},
108901 (2007).

\noindent \lbrack 29] T. Ivezi\'{c}, \textit{arXiv: 1005.3037.}

\noindent \lbrack 30] F.W. Hehl and Yu.N. Obukhov, \textit{Phys. Lett. A}
\textbf{334}, 249 (2005).

\noindent \lbrack 31] E.J. Post, \textit{Formal Structure of Electromagnetics%
} - \textit{General Covariance and }

\textit{Electromagnetics }(North-Holland, Amsterdam, 1962);

(Dover, New York, 1997).

\end{document}